\documentclass[aps,prl,twocolumn,groupedaddress]{revtex4}

\begin{document}


\title{Single-File Diffusion of Atomic and Colloidal Systems: Asymptotic Laws}


\author{Markus Kollmann}
\email[Corresponding author:]{m.kollmann@fz-juelich.de}
\affiliation{Institut f\"ur Festk\"orperforschung, Teilinstitut Weiche Materie, Forschungszentrum J\"ulich}


\date{\today}

\begin{abstract}
In this work we present a general derivation of the non-Fickian
behavior for the self-diffusion of identically interacting particle
systems with excluded mutual passage. We show that the conditional
probability distribution of finding a particle at position $x_{t}$
after time $t$, when the particle was located at $x_{0}$ at $t=0$, follows a
Gaussian distribution in the long-time limit, with variance $2W(t)\sim
t^{1/2}$ for overdamped systems and with variance $2W(t)\sim
t$ for classical systems. The asymptotic behavior of the mean-squared displacement, $W(t)$, is shown to be
independent of the nature of interactions for homogeneous systems in
the fluid state. Moreover, the long-time behavior of self-diffusion is
determined by short-time and large scale collective density fluctuations.
\end{abstract}


\maketitle


Single-file diffusion refers to the motion of particles in narrow
pores so that the individual particles are unable to pass each other
and the sequence of particle labels remains the same over time. As the
mutual passage is excluded, the motion of individual particles
requires the collective motion of many other particles in the same
direction. This restriction leads to an anomalous behavior of the
self-diffusion for overdamped systems in the long-time limit, which has been subject to
long-standing theoretical investigations \cite{Harr65,Arra83,Beij83}. For classical system normal diffusive behavior has been found for the self-diffusion function \cite{Jeps65,Lebo67}. So far, rigorous results
have been derived for the long-time behavior of the conditional
probability density function (PDF), $P(y_{t},t|y_{0},0)$, of finding a particle
at position $x_{t}$ at time $t$, being located at $x_{0}$ at time $t=0$,
 only for one-dimensional hard-rod systems using various analytical methods
 \cite{Levi73,Jeps65,Harr65,Arra83}. For these systems it has been shown that $P(y_{t},t|y_{0},0)$ is Gaussian with variance
$2 W(t)$ for asymptotically large times. The function $W(t)$ is the
mean-squared displacement (MSD) given by $W(t)=A_{o}\,t^{1/2}$ and $W(t)=A_{c}\,t$ for overdamped and classical systems, respectively. The constants $A_{o}$ and $A_{c}$ depend
 on thermodynamic properties and short-time transport coefficients of
 the particular systems. Although these results have been derived for hard-rod
 systems only, it 
is generally believed that in systems with prohibited exchange of
particle labels the MSD scales like $W(t)\sim t^{1/2}$ in overdamped systems
\cite{Alex78} and like $W(t)\sim t$ in classical systems.

Experimental evidence confirming the anomalous diffusive behavior has
been found only recently in atomic \cite{Kukl96} and colloidal systems
\cite{Wei00}. The experimental results support the conjectured
scaling behavior of the MSD as given above \cite{Alex78}. The single filing
condition is hard to fulfill experimentally for atomic systems, whereas for micrometer-sized colloidal particles narrow
channels can be easily generated \cite{Wei00}. While atoms follow ballistic motion, colloidal particles
follow overdamped motion due to the presence of the solvent. Moreover,
in colloidal systems there exist additional
correlations due to solvent mediated hydrodynamic interactions (HI)
\cite{GNhabil}. Nevertheless the experiments, using magnetic colloids with long
range dipole-dipole pair interactions, show clearly the same long-time
scaling of the MSD as predicted for hard-rod systems with HI neglected. 

So far a
general theory for the self-diffusion in homogeneous single-filing atomic and
Brownian systems was still lacking and will be given in this work. We
 show that the general conjectured scaling of
the MSD for single filing systems \cite{Alex78} is indeed true and derive asymptotic laws for the single-file diffusion of
colloidal and atomic particles systems with arbitrary interaction potential, under the condition that the correlation length between the particles is of finite range and the particles interact identically.   We prove that in limit of times, much larger
than the time needed for a particle to diffuse the mean particle
distance, the conditional PDF  $P(y_{t},t|y_{0},0)$ is normal
distributed with variance $2\,W(t)$. Moreover, we include HI in the case of colloidal systems and give explicit  
expressions for the pre-factor $A$ in the expression for the MSD for
both classical and overdamped systems.\\

{\it Overdamped systems.} We consider a one-dimensional homogeneous
system of $N$ identically interacting Brownian particles in
equilibrium. The system size is of length $L$ and assumed to be macroscopically large. The total number density of the particles in the system
is denoted by $\bar\rho=N/L$ and the mean particle distance is given by
$a={\bar \rho}^{\, -1}$. The interaction potential is chosen to be
infinite at overlap of any two particle centers. Consequently, the
particles are unable to pass each other.  In the overdamped limit the time evolution of the particle trajectories follow the stochastic equations \cite{erma78}
\begin{equation}
  \frac{\partial y_{i}(t)}{\partial t}={\hat\Omega}_{B}[y^{N}(t)]\,y_{i}(t)+\eta_{i}(t)
\label{a1}
\end{equation}
with ${\hat\Omega}_{B}$ the so-called backward Smoluchowski operator \cite{GNhabil,Risken} given by 
\begin{equation}
{\hat\Omega}_{B}[y^{N}]=\sum_{i,j=1}^{N} \left[\frac{\partial}{\partial y_{j}}-\frac{\partial U(y^{N})}{\partial y_{j}}\right]D_{ij}(y^{N})\frac{\partial}{\partial y_{i}}\quad.
\label{a1b}
\end{equation}
Here, $y_{i}(t)$ is the position of particle $i$ at time $t$,
$D_{ij}(y^{N})$ is the diffusivity tensor, accounting for solvent
mediated hydrodynamic interactions, and $U(y^{N})$ is the interaction potential
among the particles in the system, given in units of $k_{B}T$. We also make
use of the supervector notation
$y^{N}(t)=\{y_{1}(t),...,y_{N}(t)\}$. The random noise
$\eta_{i}(t)$ is normal distributed with mean zero and the second moment is given by $\langle \eta_{i}(t)\eta_{j}(t')\rangle=2D_{ij}(y^{N})\delta(t-t')$ . We emphasize that the stochastic equations have to be interpreted in the Ito sense \cite{vanKampen}.
The basic quantity we want to calculate is the MSD, $W(t)$, given by 
\begin{equation}
   W(t)=\frac{1}{2}\left\langle [y_{i}(t)-y_{i}(0)]^2 \right\rangle
\label{a2}
\end{equation}
for asymptotically large times $t$. The index $i$ is arbitrary but
chosen such that the particle under consideration is located
sufficiently far from the walls for the time-interval considered. The brackets, $\langle  \,. \,\rangle$,
indicate an average with respect to the joint probability
density function, $P(y^{N}_{t},t|y^{N}_{0},0)P_{eq}(y^{N}_{0})$, of finding the system at position
$y^{N}_{t}$ at time $t$ and at position
$y^N_{0}$ at time $t=0$ in configuration space. Here, $P_{eq}(y^{N})=\exp[-U(y^{N})]/Z_{N}$
is the Boltzmann distribution and $Z_{N}$ the configurational
integral. The conditional PDF,
$P(y^{N}_{t},t|y^{N}_{0},0)$, 
satisfies the differential equation
\begin{equation}
\partial_{t} P(y^{N}_{t},t|y^{N}_{0},0)={\hat \Omega}[y^{N}_{t}]\, P(y^{N}_{t},t|y^{N}_{0},0)
\label{a2b}
\end{equation}
with initial condition 
\begin{equation}
\lim_{t \rightarrow 0^{+}}
P(y^{N}_{t},t|y^{N}_{0},0)=\delta(y^{N}_{t}-y^{N}_{0}).
\label{a2c}
\end{equation}
In Eq. (\ref{a2b}), the operator $\hat\Omega$ is the Hermitian
conjugate of the operator  $\hat\Omega_{B}$.
As we assume the system to be ergodic and in equilibrium, the
conditional PDF $P(y^{N}_{t},t|y^{N}_{0},0)$
will approach the Boltzmann distribution, $P_{eq}(y^{N})$, in the long-time limit. Hence, $\lim_{t\rightarrow \infty}W(t)$ is of quadratic order in the
system size. As $W(t)$ is a monotonically increasing function of time,
the major contribution to the MSD at large times is given by particle
trajectories whose end to end distances are much larger than the mean
particle distance, $a$. In the following we look for an expression for
the MSD in the limit $t \gg \tau=a^2/D_{S}$,
with $D_{S}$ the short-time self-diffusion coefficient of a free
diffusing particle. 

To simplify the calculation for $W(t)$ we introduce an auxiliary stochastic function $x(t)$, defined by $y_{i}(t)<x(t)<y_{i+1}(t)$, and the microscopic density $\hat\rho(y,t)=\sum_{i}^{N}\delta(y-y_{i}(t))$. By definition we have $W(t)={\tilde W}(t)+{\cal O}(a)$ with ${\tilde W}(t)=1/2\left\langle [x(t)-x(0)]^2 \right\rangle$. Consequently, the long-time behavior of $W(t)$ and ${\tilde W}(t)$ for infinite system size are related by 
\begin{equation}
\lim_{t\rightarrow \infty}\lim_{L\rightarrow \infty}\frac{{\tilde W}(t)}{W(t)}=1 \quad.
\label{a3}
\end{equation}
In order to calculate ${\tilde W}(t)$ we have to derive an explicit expression for the conditional PDF, ${\tilde P}(x_{t},t|x_{0},0)$, of finding trajectory of $x(t)$ at position $x_{t}$ at time $t$ when it was located at $x_{0}$ at time $t=0$.

To this end, it is useful to introduce another stochastic variable, $y_{i+M}(t)<z(t)<y_{i+M+1}(t)$, 
and take $M$ so large such that $\left\langle x(t')z(t'') \right\rangle=0$ for all times $t',t''\leq t$. By homogeneity of the systems, $x(t)$ and $z(t)$ are stochastically equivalent variables. Now, the conditional PDF, ${\tilde P}_{r}(x_{t}-z_{t},t|x_{0}-z_{0},0)$, to find the two trajectories, $x(t)$ and $z(t)$, separated a distance $x_{t}-z_{t}$ at time $t$ when they were separated a distance $x_{0}-z_{0}$ at time $t=0$ is then given by
\begin{eqnarray}
{\tilde P}_{r}(\Delta x_{t}-\Delta z_{t},t)&=&\int_{L} {\tilde P}(\Delta x_{t}+a,t){\tilde P}(\Delta z_{t}+a,t)da\nonumber\\
\quad\label{a3b}
\end{eqnarray}
Here, $\Delta x_{t}=x_{t}-x_{0}$, $\Delta z_{t}=z_{t}-z_{0}$ and we employed translationary invariance and stationarity for the conditional PDF, ${\tilde P}(\Delta x_{t},t)={\tilde P}(x_{t},t|x_{0},0)$. From the knowledge of ${\tilde P}_{r}(\Delta x_{t}-\Delta z_{t},t)$, one can determine uniquely the conditional PDF ${\tilde P}(\Delta x_{t},t)$ by making use of the convolution theorem.

In the following we are going to determine the function ${\tilde P}_{r}(\Delta x_{t}-\Delta z_{t},t)$ from the stochastic properties of the microscopic density, $\hat\rho(y,t)$. To this end, we introduce the functional 
\begin{equation}
h(x_{t},z_{t}|x_{0},z_{0}):= \int_{z_{t}}^{x_{t}}\hat\rho(y,t)\,dy- \int_{z_{0}}^{x_{0}}\hat\rho(y',0)\,dy'\,,
\label{a4}
\end{equation}
together with the general condition
\begin{equation}
\lim_{t\rightarrow\infty}\lim_{N,L\rightarrow\infty}\mbox{Prob}\left\{(x(t)-x(0))^2>{\tilde W}(t)^{1+\epsilon}\right\}=0\,,
\label{a4b}
\end{equation}
with $\epsilon>0$ arbitrary small.
Consequently, we get by definition of $x(t)$, $y(t)$
\begin{equation}
h(x(t),z(t)|x(0),z(0))=0\,. 
\label{a5}
\end{equation}
The conditional PDF, ${\tilde P}_{r}(\Delta x_{t}-\Delta z_{t},t)$, is then given in long-time regime by  
\begin{eqnarray}
{\tilde P}_{r}(\Delta x_{t}-\Delta z_{t},t)&=&\lim_{l\rightarrow\infty}\lim_{t \gg \tau}\lim_{N,L\rightarrow\infty}\bigg\langle \delta_{l}\big[h(x_{t},z_{t}|x_{0},z_{0})\big]\nonumber\\
&&\qquad\qquad  *\bigg| \frac{\partial h(x_{t},z_{t}|x_{0},z_{0})}{\partial(x_{t}-z_{t})}\bigg| \bigg\rangle\,.
\label{a6}
\end{eqnarray}
Here, $t \gg \tau$ indicates the asymptotic limit for large times and
$\delta[z]=\lim_{l\rightarrow\infty}\delta_{l}[z]$ is the delta distribution in the Dirac sense. The reason why we take the
asymptotic limit $t\gg\tau$ before taking the limit $l\rightarrow\infty$ is
because the conditional probability density, ${\tilde P}_{r}(\Delta x_{t}-\Delta z_{t},t)$, can not be normalized in a simple way if $l\rightarrow\infty$ is      
taken first. By definition of $x(t)$ there
corresponds an infinite set of realizations of ${\hat\rho}(y,t)$ to one
given realization of $x(t)$ which leads to divergent
contributions for $\left\langle\delta\left[h(x_{t},z_{t}|x_{0},z_{0})\right] \right\rangle$. Performing
the asymptotic analysis for long times 
first, we expect to loose all informations on scales of the mean
particle distance and hence there is some chance that the same form
of normalization is valid, as used in Eq. (\ref{a6}) for a finite set of realizations of ${\hat\rho}(y,t)$ corresponding to one realization of $x(t)$. 
We emphasize that the right-hand-sides of Eq. (\ref{a3b}) and Eq. (\ref{a6}) are only stochastical equivalent in the long-time limit, as in Eq. ({\ref{a6}) we get additional fluctuations because in Eq. (\ref{a4}) there enters no information that the amount of trajectories of $x(t)$ traveling infinitely far during a finite time is of measure zero. This additional fluctuations do not contribute in the long-time limit, as can be shown by inserting in Eq. (\ref{a9}) any value between the upper bound, as given in Eq. (\ref{a4b}), and the lower bound, $x(t)-x(0)=0$, for $x(t)$.

The conditional PDF for the relative motion, ${\tilde P}_{r}(\Delta x_{t}-\Delta x_{0},t)$, can be rewritten in integral form 
\begin{eqnarray}
{\tilde P}_{r}(\Delta x_{t}-\Delta z_{t},t)&=&\lim_{l\rightarrow\infty}\lim_{t \gg
  \tau_{I}}\int\Bigg|\frac{1}{2\pi\,i\,\xi}\frac{\partial}{\partial(x_{t}-z_{t})}\nonumber\\   
&&*\exp\left[\sum_{n=1}^{\infty}\frac{(i\,\xi)^{n}}{n!}\kappa^{(n)}-\xi^{2}/(2l)\right]d\xi\Bigg|\nonumber\\
\label{a7}
\end{eqnarray}
with $\kappa^{(n)}$ the $n$-th cumulant of $h(x_{t},z_{t}|x_{0},z_{0})$.
In the following we show that in the limit $t\gg \tau$ only the first two cumulants contribute to ${\tilde P}_{r}(\Delta x_{t}-\Delta z_{t},t)$. 
The first cumulant reads simply 
\begin{equation}
\kappa^{(1)}=\langle h(x_{t},z_{t}|x_{0},z_{0}) \rangle={\bar \rho}\,[(x_{t}-z_{t})-(x_{0}-z_{0})]\,.
\label{a8}
\end{equation}

The second cumulant can be expressed as 
\begin{widetext}
\begin{eqnarray}
\kappa^{(2)}&=&\left\langle h(x_{t},z_{t}|x_{0},z_{0})^2\right\rangle-\left\langle h(x_{t},z_{t}|x_{0},z_{0}) \right\rangle^2 \nonumber\\
&=&\frac{\bar\rho}{\pi}\int\, \frac{1}{q^{2}}\Big\{2\,S(q,0)-\big(\exp[-i\,q\,(x_{t}-x_{0})]+\exp[i\,q\,(z_{t}-z_{0})]\big)S(q,t)\Big\}\,dq 
\label{a9}
\end{eqnarray}
\end{widetext}
Here, we introduced the dynamic structure factor defined by
$S(q,t)=\left\langle \delta\hat\rho(q,t)\delta\hat\rho^{*}(q,0)
\right\rangle/N$, with $\delta\hat\rho(q,t)=\int dy\,
\exp[i\,q\,y](\hat\rho(y,t)-\bar\rho)$ and made use of translationary invariance of the system.
By inspection of  Eq. (\ref{a9}) we find that the dominating contributions to
the integral at large times come from values $q \ll a^{-1}$. 

To evaluate $S(q,t)$ in the limit of small values of $q$, we
make use of the Mori-Zwanzig projector operator formalism
\cite{Mori65,Berne}. For this purpose we introduce the projection operators
$\hat P=|\hat\rho(q,0)\rangle_{eq}\langle\hat\rho(q,0)|/(N\,S(q,0))$
and ${\hat Q}=1-\hat P$. Here, we make use of the Dirac notation and
the brackets,
 $\langle\,.\, \rangle_{eq}$, indicate averaging with respect
to Boltzmann weight. The time derivative of the dynamic structure
factor can be written within this formalism as
\begin{equation}
\frac{\partial S(q,t)}{\partial t}= -q^2\,D^{e}(q)\,S(q,t)+\int_{0}^{t}M(q,t-t')\,\frac{S(q,t')}{S(q,0)}\,dt'
\label{a9b}
\end{equation}
with an effective wavevector dependent diffusion function,
 $D^{e}(q)=H(q)/S(q)$ and the so-called memory function
 $M(q,t)=\langle \hat \rho(q,0)\,\Omega_{B}|\,Q\,\exp[Q\Omega_{B}Q\,t]
\,Q\,|\Omega_{B}\,\hat \rho(q,0) \rangle_{eq}$
 \cite{GNhabil}. Here, $H(q)=\langle\hat \rho(q,0)|\Omega_{B}|\hat
\rho(q,0)\rangle_{eq}/N$ is the hydrodynamic function \cite{Dhont}. Detailed calculations show that $M(q,t)={\cal
 O}(q^4)$ if HI is 
neglected, or taken into account in far-field approximation, 
$D_{ij}(y^N)=D_{ij}(y_{i}-y_{j})$ \cite{GNhabil}. The
full treatment of HI in the memory function gives contributions of
order $q^2$, but the correction to $S(q,t)$ are
only within a few percent, even for highly concentrated suspension
with short-range interactions \cite{Szymczak}. For the colloidal systems used in the
experiments of Ref. \cite{Wei00}, the far-field approximation of $D_{ij}(y^N)$ has been shown to be an excellent one \cite{Rinn99}.   

For identical Brownian particles and HI neglected or treated on pairwise additive level the solution of Eq. (\ref{a9b}) for $S(q,t)$ is given by
\begin{equation}
S(q,t)=S(q,0)\,\exp[-q^2 D^{e} t]+{\cal O}(q^{4}t)\quad,
\label{a10}
\end{equation}
with $D^{e}=\lim_{q\rightarrow 0}D^{e}(q)$. Substituting
Eq. (\ref{a10}) into Eq. (\ref{a9}), the second cumulant reads, in the
asymptotic long-time regime, with $S=S(0,0)$,
\begin{equation}
\kappa^{(2)}=4\,\bar\rho\, S\,\left(\frac{ D^{e} t}{\pi}\right)^{1/2}+{o}(t^{\epsilon/2}) \quad.
\label{a11}
\end{equation}
Here, the dependence of $\kappa^{(2)}$ on $x_{t}-x_{0}$ is of order ${o}(t^{\epsilon})$ due to the condition Eq. (\ref{a4b}).
The higher cumulants are given by $\kappa^{(2n+1)}=0$ and $\kappa^{(2n)}={\cal O}(t^{1/2})$ for $n\in\{1,2,...\}$.
In order to prove that the odd cumulants higher than one are exactly zero, we have
made use of the backward operator to be Hermitian with respect to the
Boltzmann
weighted inner product, $\langle f\,\Omega_{B}\,g \rangle_{eq}=\langle
g\,\Omega_{B}\,f \rangle_{eq}$, and invariance of the
system with respect to translation and space inversion. 
The asymptotic scaling of the even cumulants can be found with
the help of the mean value theorem of integration and the fact that
the correlation length between particles can be taken finite for physical
systems in the fluid state. For a $2n$-point correlation function one
uses the mean-value theorem for $2n-2$ integrals and then proceeds
along the same line as for the second cumulant.

Using the explicit expressions for the cumulants we find from Eq. (\ref{a7})  
\begin{eqnarray}
{\tilde P}_{r}(\Delta x_{t}-\Delta z_{t},t)&=&\lim_{l\rightarrow\infty}\frac{\bar\rho}{2\pi}\int d\xi\, \exp\left[\sum_{n=1}^{\infty}\frac{(i\,\xi)^{n}}{n!}\kappa^{(n)}\right.\nonumber\\&&\qquad\qquad\qquad\qquad\qquad-\xi^{2}/(2l)\Bigg]\nonumber\\
&=&\hspace{-1mm}\frac{(\bar\rho/(2\,S))^{1/2}}{2\,(\pi D^{e}t)^{1/4}}\exp\left[-\frac{\bar\rho\,(\Delta x_{t}- \Delta z_{t})^2}{8\,S\,(D^{e}t/\pi)^{1/2}}\right]\nonumber\\
\label{a13}
\end{eqnarray}
From Eq. (\ref{a3b}) we can derive finally an expression for the conditional PDF for $x(t)$, 
\begin{equation}
{\tilde P}(x_{t},t|x_{0},0)=\frac{\bar\rho^{\,1/2}}{2\,S^{\,1/2}(\pi D^{e}t)^{1/4}}\exp\left[-\frac{\bar\rho\,(x_{t}- x_{0})^2}{4\,S\,(D^{e}t/\pi)^{1/2}}\right].\\\
\end{equation} 
Hence, the mean-squared displacement in the asymptotic limit is
given by 
\begin{equation}
W(t)=\frac{S}{\bar\rho}(D^{e}t/\pi)^{1/2}+o(t^{\epsilon/2}) \qquad \mbox{for} \quad t\gg \tau
\label{a14}
\end {equation}
The effective diffusion constant $D^{e}$ can be experimentally
determined by a short-time measurement of $S(q,t)$ at $q\ll
a^{-1}$ \cite{Clemens}. For times, short enough that each individual particle does not 'feel' the presence of the other particles by direct interactions, the integral over the memory function in
Eq. (\ref{a9b}) does not contribute and thus there exists a unique
relationship between $S(q,t)$ and $D^{e}$ as given in
Eq. (\ref{a10}) with the second term on the right-hand-side
neglected. It is interesting to see that a long-time self-diffusion
property like the MSQ can be determined to high accuracy from the
short-time collective behavior of the system. This fact is a direct
consequence of the single-filing condition imposed on our system.\\
{\it Classical Systems.} So far we have considered only interacting Brownian particles in the
overdamped limit. For classical particle systems following Liouville
dynamics, like atoms or molecules in the high temperature limit, we can
essentially follow the same route as for the overdamped systems with
the Liouville operator 
substituted for the Smoluchowsky operator. 
The analysis follows the same line as for overdamped systems with the result that the conditional PDF for classical systems is again Gaussian to leading order in time but normal diffusive,
\begin{equation}
P(x_{t},t|x_{0},0)=(4\pi\,D_{s}\,t)^{-1/2}\exp\left[-\frac{(x_{t}-x_{0})^2}{4\,D_{s}\,t}\right]\quad,
\end{equation}
with self-diffusion constant $D_{s}={\bar v}\,(1-d\bar\rho)/(2\,\bar\rho)$ for hard-rod systems (c.f. Ref(\cite{Jeps65,Lebo67})) and $D_{s}=c_{s}\,S\,C_{V}/(2\,C_{P}\,\bar\rho)^{-1}$ for classical systems in the hydrodynamic regime \cite{note3}. Here, $\bar v$ is the average velocity of the particles, $d$ is the rod length, $c_{s}$ is the speed of sound and $C_{V}$, $C_{P}$ are the specific heats at constant volume and pressure, respectively. If the low-frequency sound modes are suppressed, like in the presence of a randomizing background (e.g. porous media), we expect the MSD to scale like $W(t)\sim t^{1/2}$ \cite{Levi73}. This interesting property of single-filing systems will be subject of a forthcoming paper.
In conclusion we have derived a general theory for the asymptotic behavior of the MSD in single-filing systems. We have shown that $W(t)$ is determined by the short-time and large-scale collective behavior of the particles in the system. This relation holds true for any kind of interaction potential as
long as correlation length between the particles is finite, which is
true for any physical system in the fluid state. Moreover, this relation is unique in the case of classical systems and
in the case of overdamped systems with HI taken into account in far-field
approximation.

\begin{acknowledgments}
We thank C. Bechinger, J.K.G Dhont and G. N\"agele for helpful discussions.
\end{acknowledgments}

\bibliography{sfd.bib}

\begin{thebibliography}{20}
\bibitem[{not()}]{note3}
\bibinfo{note}{Hard-rod systems are special in the sense that they do not
  'thermalize' locally. Thus it makes no sense to speak about a sound wave in
  such a system.}
\end{thebibliography}

\end{document}